\documentstyle[festkoer,psfig]{vieweg}

\author{M.~Ulmke$^1$, P.~J.~H.~Denteneer$^2$, V.~Jani\v s$^3$,
R.~T.~Scalettar$^4$, A.~Singh$^{1,*}$, 
D.~Vollhardt$^1$, G.~T.~Zimanyi$^4$}

\Kurzautor{Ulmke et al.}

\title{Disorder and Impurities in Hubbard-Antiferromagnets}

\Kurztitel{Disorder and Impurities in Hubbard-Antiferromagnets}

\Adresse{$^1$Theoretische Physik III, Elektronische Korrelationen 
und Magnetismus, Institut f\"ur Physik,                     
 Universit\"at Augsburg, 86135 Augsburg,
Germany\\
$^2$Lorentz Institute for Theoretical Physics, 
University of Leiden, P.O.~Box 9506, 2300 RA Leiden, The Netherlands\\
$^3$Institute of Physics, Academy of Sciences of the Czech Republik, CZ 18040
Praha 8, Czech Republik\\
$^4$Department of Physics, University of California, Davis, CA 95616, USA}

\begin{document}

\Titel

\begin{abstract}
We study the influence of disorder and randomly distributed 
impurities on the properties of correlated antiferromagnets.
To this end the Hubbard model with (i) random potentials,
(ii) random hopping elements, 
and (iii) randomly distributed values of interaction
is treated using quantum Monte Carlo and dynamical mean-field theory.
In cases (i) and (iii) weak disorder can lead to an enhancement
of antiferromagnetic (AF) order: 
in case (i) by a disorder-induced delocalization,
in case (iii) by binding of free carriers at the impurities. 
For strong disorder or large impurity concentration antiferromagnetism
is eventually destroyed.
Random hopping leaves the local moment stable but AF order 
is suppressed by local singlet formation.
Random potentials induce impurity states within the charge gap until
it eventually closes.
Impurities with weak interaction values shift the Hubbard gap
to a density off half-filling. In both cases an antiferromagnetic phase
without charge gap is observed.

\end{abstract}

\section{Introduction}

Antiferromagnetic spin correlations are present in many strongly correlated
electron systems, notably the prototype Mott insulators NiO and $\rm V_2O_3$,
the parent compounds of HTSC cuprates, and heavy fermion systems such
as YbP, $\rm U_2Zn_{17}$, and many others.
Many of those systems are intrinsically disordered, in particular upon
additional 
homo- or heterovalent 
doping. The influence of impurity doping
on antiferromagnetic (AF) order and electronic properties 
has recently been studied in a variety of systems. 
Doping with static scatterers like nonmagnetic impurities
usually weakens antiferromagnetic order, a prominent
example being Zn doping in $\rm Y Ba_2 Cu_3 O_6$ \cite{ZN1}. 
In spin chains 
($\rm CuGeO_3$) \cite{Hase} and ladder compounds ($\rm SrCu_2O_3$) 
\cite{ZN2} on the other hand, 
doping with magnetic and nonmagnetic impurities 
can \em induce \em AF order while the pure systems show spin gap behavior.
Very effective
in destroying AF order are mobile carriers, 
e.g.~hole doping in $\rm La_{1-x}Sr_xCuO_4$ \cite{LaSrCuO}. 
The stability of AF order strongly depends
on the positions of the dopant level. While in the nickel oxides
$\rm La_{1-x}Sr_xNiO_4$
 \cite{LaSrNiO} and $\rm Ni_{1-x}Li_x O$ \cite{Huefner} holes are 
supposed to be localized, in the cuprate $\rm La_{1-x}Sr_xCuO_4$ the hole level
lies in the valence band leading to mobile scatterers. As a result,
AF order is stable in $\rm La_{1-x}Sr_xNiO_4$ up to $x=0.5$, but is 
destroyed in the cuprate already at approx.~5\% Sr doping.

In the present paper we will study the influence of disorder
on AF order and the Mott band gap in correlated antiferromagnets.
We employ the Hubbard model in the presence of different types
of disorder. While the (disordered) Hubbard model is certainly far too simple
to describe real materials it already contains very rich physics
including local moment formation, magnetic ordering,
Mott-Hubbard transition, and Anderson localization.
On the other hand, the interplay of disorder and interactions
in electronic systems belongs to the most difficult problems
in physics, and reliable results 
within simple models are still very desirable.
The problem has been investigated in the past by a variety of methods,
including field theoretical approaches \cite{mafrad}, 
renormalization group treatments \cite{fink2,ma}, 
unrestricted Hartree-Fock \cite{logan,uhf}, 
dynamical mean-field theory (DMFT) \cite{juv,dk,ulmke}, 
quantum Monte Carlo (QMC) \cite{sandvik1,ulmke2,ulmke3}, 
and several more (see \cite{belitz} for a review). 
Here we give an overview of results obtained
by QMC and DMFT concentrating on the AF phase diagram and the 
Mott gap.

We consider the following Hubbard Hamiltonian:
\begin{equation} 
\hat H = \sum_{i\sigma} (\epsilon_i-\mu) \hat{n}_{i\sigma}
 + \sum_{\langle ij \rangle \sigma} t_{ij} 
(\hat{c}_{i \sigma}^{\dagger} \hat{c}_{j \sigma} + {\rm h.c.}) +
 \sum_{i} U_i (\hat{n}_{i \uparrow}-\frac{1}{2}) 
(\hat{n}_{i \downarrow} -\frac{1}{2}) .
\label{Eq:dishub}
\end{equation}
In principle all parameters $\epsilon_i, t_{ij}, U_i$
can be randomly distributed. The precise definition of the
different disorder types studied in this paper will be given in the
following sections. 
The average $t\equiv \langle t_{ij}\rangle$ sets our energy scale.
We will restrict the hopping $t_{ij}$ to nearest-%
neighbors hence not allowing for frustration. Longer range, random hopping
amplitudes will be important in the modeling of amorphous materials
such as doped semiconductors \cite{belitz,bhatt} 
and are not considered in the present work.

\section{Methods}

\subsection{Determinant quantum Monte Carlo (d=2)}

We use a finite temperature determinant quantum Monte Carlo method 
\cite{blankenbecler}
to obtain approxi\-mation-free results for finite lattices.
The algorithm is based on a mapping of the interacting
electron problem onto a $d+1$ dimensional quasi-classical
problem using auxiliary Ising-type spins.
It provides for calculating thermal averages of observables, $\hat A$,
at a temperature $T=1/\beta$,
\begin{equation} 
\langle \hat A\rangle = 
\frac{{\rm Tr\,}\hat A\, e^{-\beta \hat H} }{{\rm Tr\,} e^{-\beta \hat H}} .
\end{equation} 
The phase space sampling over the auxiliary field configurations
is performed using Monte Carlo techniques. The weight of 
a configuration is proportional to a product of two determinants,
one for each electron spin species. 
In the case of half-filling without random potentials, 
i.e.~$\epsilon_i-\mu\equiv 0$, on a bipartite lattice
the determinants always have the same sign, hence their product
is always positive semi-definite, which can be shown
by  particle-hole transformation of one spin species
$[c_{{\bf i}\downarrow} \rightarrow 
(-1)^{{\bf i}} c_{{\bf i}\downarrow}^\dagger]$.
In the general situation that the product can become negative 
the algorithm can still be employed in principle. However, the 
signal to noise ratio decreases exponentially with systems size,
inverse temperature, and interaction, putting severe restrictions to 
the applicability of the method.
This so-called ``minus-sign problem'' is a general obstacle
for all exact fermionic Monte Carlo methods as well as for spin-systems
in the presence of frustration.
Even without the minus-sign problem the computational effort is
large because the computer time grows cubically with system size $N$,
restricting $N$ to the order of 100 on present supercomputers.

In the case of disorder all observables have to be averaged over
the (frozen) disorder configurations. 
Because of the computational effort we restrict ourselves to
two dimensional lattices with linear size up to $L_x=10$
which often allows a reliable finite size scaling.
Since we are interested in AF ordering we calculate the magnetic 
correlation functions $C({\bf l})$ and their Fourier transforms,
the magnetic structure factors $S({\bf q})$, 
\begin{equation}
C({\bf l})  = 
\frac{1}{N} \sum_{{\bf j}} \langle m_{\bf j} m_{{\bf j+l}} \rangle ,
\quad
S({\bf q}) = \sum_{l} C({\bf l}) e^{i {\bf ql}} \label{strucq} .
\label{correl}
\end{equation}

In particular the AF structure factor $S(\pi,\pi)$ is used to obtain
the ground state sublattice magnetization $M$ by a finite size scaling
Ansatz according to spinwave theory \cite{huse}:
\begin{equation}
\frac{S(\pi,\pi)}{N} = \frac{M^2}{3} + {\cal O} (\frac{1}{L_x}).
\label{Eq:fss}
\end{equation}

\subsection{Dynamical mean-field theory (limit of infinite dimensions)}

The dynamical mean-field theory \cite{metzner,pruschke} 
is a local approximation 
in which the self energy becomes site diagonal, or momentum independent: 
\begin{equation}
\Sigma_{ij}(\omega) = \delta_{ij} \Sigma(\omega) , \qquad
\tilde\Sigma({\bf k},\omega) = \tilde\Sigma(\omega) .
\end{equation}

The one-particle Green function $G({\bf k},\omega)$ can hence be 
obtained from the non-interaction Green function $G^0({\bf k},\omega)$ by
$G({\bf k},\omega) = G^0({\bf k},\omega-\tilde\Sigma(\omega))$,
and the local Green function is given by
$G_{ii}(\omega) = 1/N \sum_{\bf k}G({\bf k},\omega) $
This does not imply a simple shift of energies, like in 
traditional mean-field theories (e.g.~Hartree-Fock), because
$\Sigma$ remains  dynamical, i.e.~frequency dependent, 
preserving local quantum fluctuations.
The local approximation becomes exact in the limit of infinite spatial
dimensionality and maps the interacting lattice model onto a self-consistent
single impurity model like, for example, the Wolff model \cite{wolff}:
\begin{equation} 
\hat H_{\rm Wolff}= 
\sum_{{\bf k}\sigma} \varepsilon_{\bf k}  \hat{n}_{\bf k}
+  U (\hat{n}_{0 \uparrow}-\frac{1}{2}) 
(\hat{n}_{0 \downarrow} -\frac{1}{2}) + 
\epsilon (\hat{n}_{0 \uparrow}+ \hat{n}_{0 \downarrow}) .  
\label{wolff}
\end{equation} 

Here the one-particle energies $ \varepsilon_{\bf k} $ have to be defined 
such that the non-inter\-acting local Green function of the Wolff model
fulfills $G^0_{\rm Wolff}\equiv (G_{ii}-\tilde\Sigma)^{-1}$.
In the self-consistent solution the local (interacting) Green function of
the Wolff model has to be equal to $G_{ii}$.
In the presence of disorder one has to average over all
possible values of $U$ or $\epsilon$, respectively.
This type of local averaging is equivalent to the ``coherent
potential approximation'', well known from investigations of 
disordered alloys. 

While the self-consistency is rather easily reached by iteration
the solution of the single impurity problem is the hard part.
There exists no analytic solution and different numerical and
approximative techniques have been employed 
(see, e.g., \cite{pruschke}). 
Here we again use
auxiliary field QMC \cite{hirsch}, an algorithm quite similar
to the one for finite dimensional lattices sketched above.
The computer time grows
like $L^3$ where the number of Matsubara frequencies, $L \propto \beta$.
Fortunately, QMC for the single band model is free from the minus-sign
problem.

In the following the non-interacting DOS is chosen as a semi-elliptic 
model DOS with bandwidth=8, equal to the $d=2$ tight binding bandwidth 
for $t=1$.
A typical quantity under consideration is the staggered magnetic
susceptibility $\chi_{\rm AF}$
whose divergence signals the transition to an AF ordered
state. 
One can also extend the DMFT equations to the ordered phase to obtain
spin and sublattice dependent electron densities and the 
sublattice magnetization $M$.
The one-particle density of states (DOS) is obtained by analytical
continuation of the imaginary time Green function using the 
Maximum Entropy method.
For details of the algorithm, the implementation of disorder averages
and determination of expectation values see \cite{pruschke,ulmke}.

\section{Random potentials}

\subsection{Local moment quenching}

\label{Sec:localmoments}
Random potentials are the most frequently studied type of disorder
in the context of Anderson localization.
Contrary to the Hubbard interaction which at half-filling favors single 
occupation on each site, different local potentials lead to 
different occupations and 
hence to a quenching of local magnetic moments on sites
with large absolute value of the local potential.
This is seen in Fig.~1a where the average local moment squared, 
$m^2$, is plotted
versus disorder strength for a flat distribution of $\epsilon$ values
with width $\Delta$. It is also shown that the spin-spin-correlations
go in parallel with $m^2$. For a large width of random potentials this
type of disorder is apparently very effective in the destruction of 
magnetic order.
To study the behavior of the charge gap, the electronic compressibility, 
$\kappa\equiv\partial n/\partial \mu$, is calculated as a function of $\Delta$
(Fig.~1b).
While disorder decreases $\kappa$ in the non-interacting case,
$\kappa$ is enhanced by disorder at finite $U$. The reason is the 
introduction of states within the AF charge gap as will be discussed below. 
\begin{figure}[b]
\vspace*{-5mm}
\psfig{file=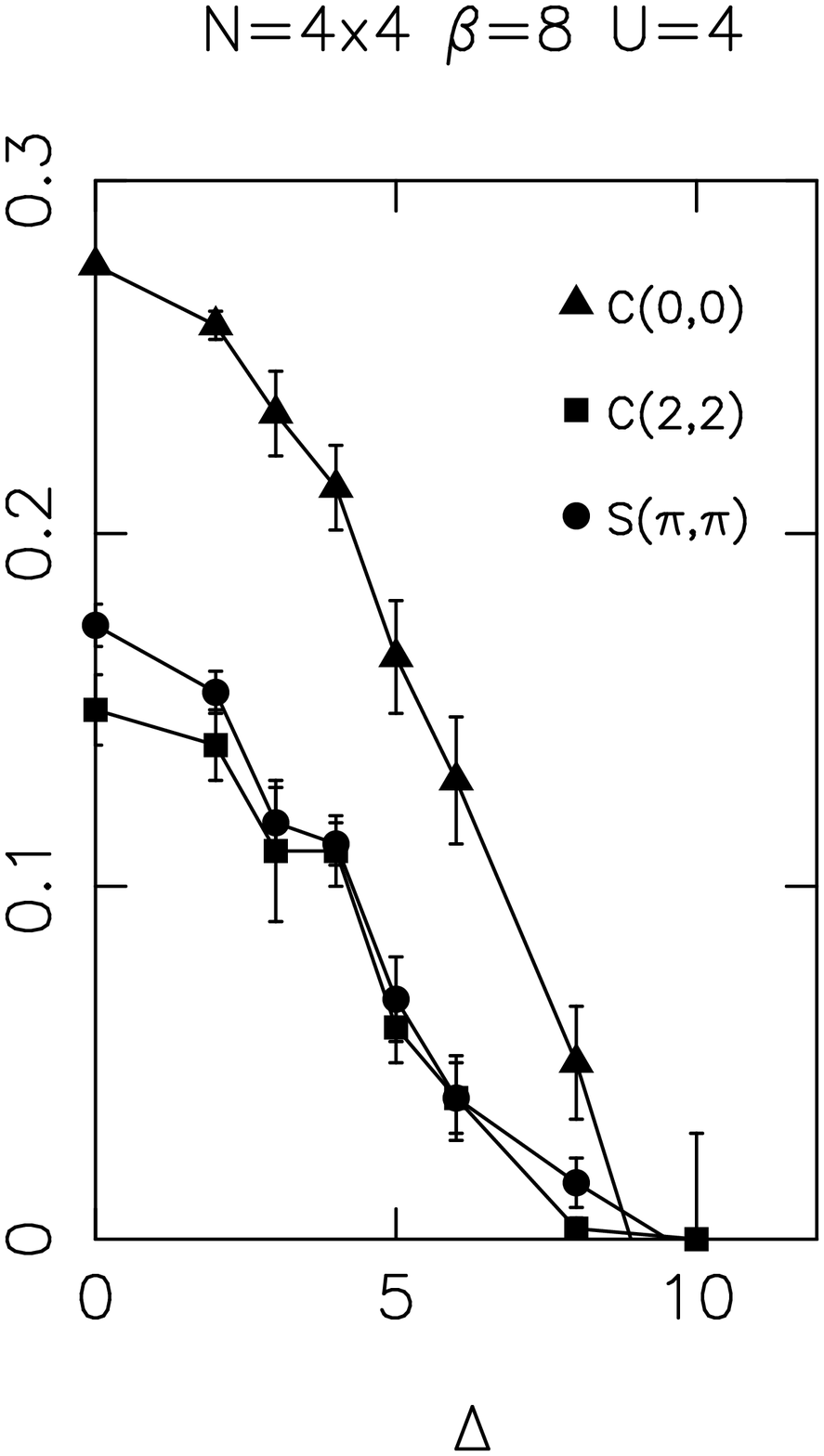,height=0.65\hsize,width=0.6\hsize}
\vspace*{-79.5mm}
\hspace*{60mm}
\psfig{file=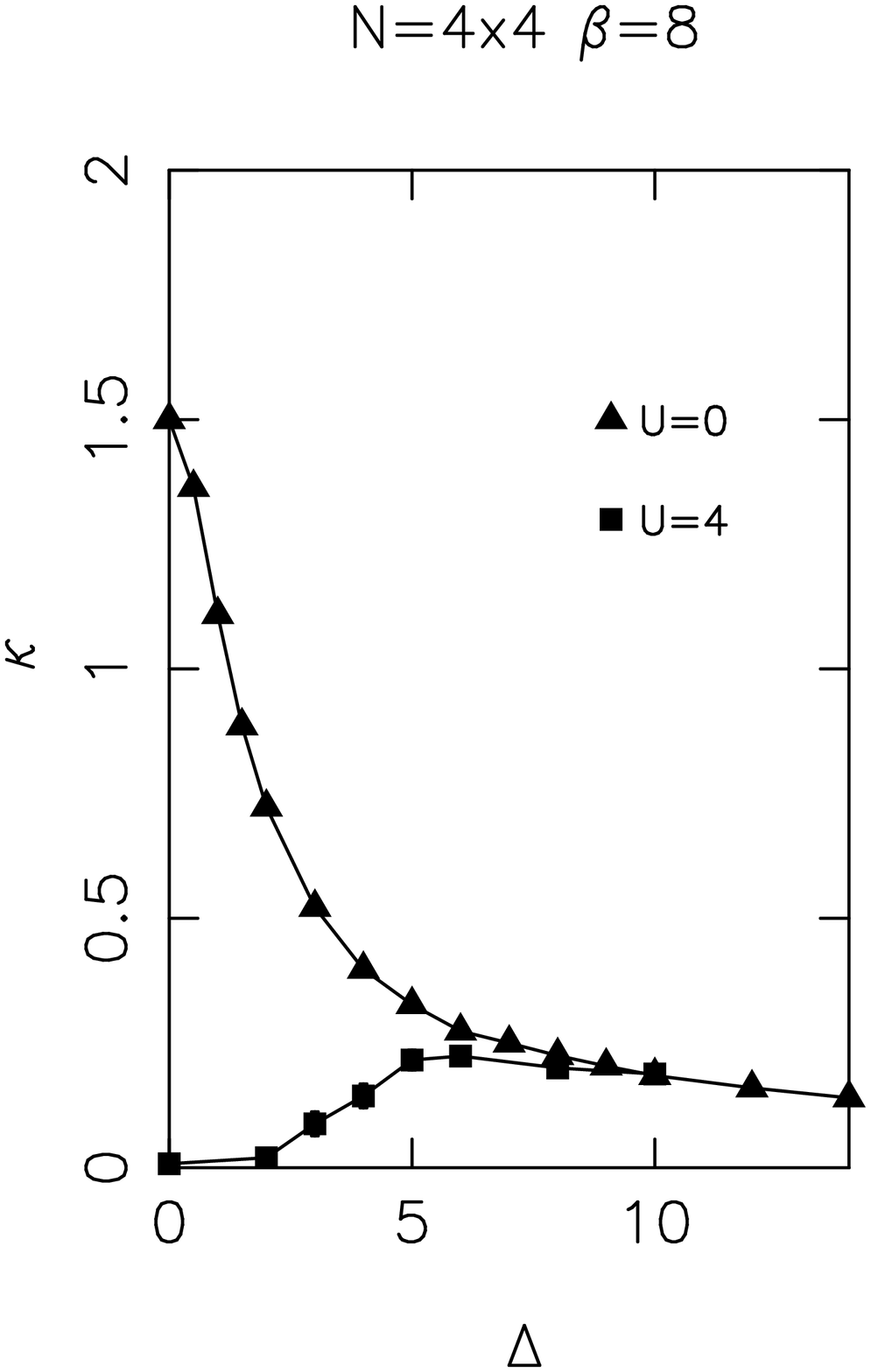,height=0.65\hsize,width=0.6\hsize}
\vspace*{-10mm}
    \par\makebox[0.4\hsize]{\small (a)}\hspace{\fill}%
    \makebox[0.55\hsize]{\small (b)}
\caption{(a) Local magnetic moment $C(0,0)$, spin-spin correlation function
at largest possible distance $(2,2)$, and AF structure factor $S(\pi,\pi)$
as a function of disorder strength for a constant distribution
of random potentials (in all cases the values at $U=\Delta=0$  are subtracted).
(b) compressibility $\kappa$ vs.~$\Delta$; disorder reduces $\kappa$ 
in the non-interacting case and enhances it at $U=4$ \cite{ulmke2}. 
}
\end{figure}
As mentioned above, random potentials break the particle-hole
symmetry leading to a minus-sign problem even at half filling.
This is the reason why only small lattices ($4\times 4$ in Fig.~1)
can be studied for this type of disorder.

\subsection{Disorder-enhanced delocalization}
Consider the situation deep in the AF phase, i.e.~with a
large staggered moment.
A high value of the local potential on a given site 
reduces the potential barrier for a majority spin electron
to tunnel to a neighboring sites. The electron is hence 
delocalized and the local magnetic moment is reduced.
A low potential on the other hand cannot significantly further
localize the majority spin because it is already almost
saturated.
This asymmetry of the localizing and delocalizing effects of 
random potentials is depicted in the DMFT results, Fig.~2.
The majority spin $(\uparrow)$ is strongly reduced for $\epsilon>0$
but almost unchanged for $\epsilon<0$. Note that the total local density
monotonically decreases with $\epsilon$, and that the net magnetization
$(n_\uparrow-n_\downarrow)$ is reduced for any $|\epsilon|>0$.

\begin{figure}[thb]
\vspace{-10mm}
\psfig{file=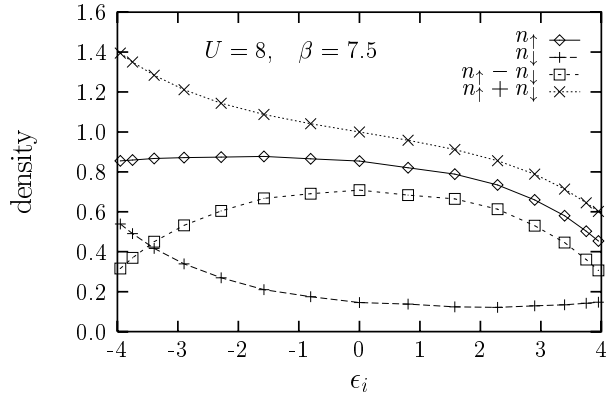,height=0.6\hsize}
\vspace{-5mm}
\caption{
Local spin resolved densities in DMFT as a function 
of the random potential value $\epsilon_i$ \cite{singh98}.
Width of disorder distribution is $\Delta=8$(=bandwidth).
}
\end{figure}
\begin{figure}
\vspace{-10mm}
\hspace*{15mm}
\psfig{file=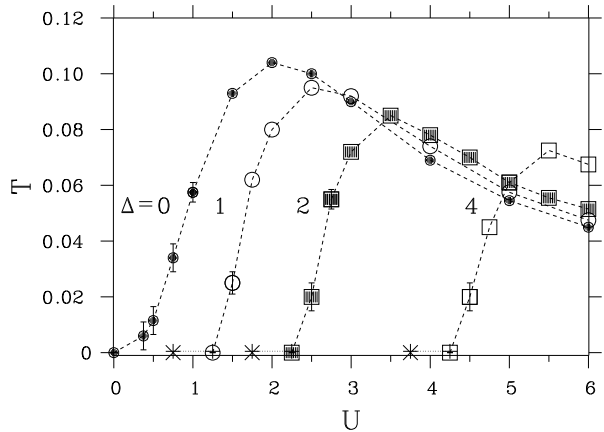,height=0.6\hsize}
\vspace{-5mm}
\caption{
AF ($T-U$) phase diagram in DMFT for a bimodal 
random potential distribution. Energies in units of the half bandwidth.
\cite{ulmke}}
\end{figure}

The fact that the delocalization is not compensated 
can be expressed in terms of an enhanced effective hopping
parameter $t_{\rm eff}$.
In the case $(U-\Delta)\gg t$ the effective $t_{\rm eff}$ can 
be estimated by the mapping onto the AF Heisenberg model: 
the AF exchange arises from virtual hopping of an electron
with spin $\sigma$ to a neighboring site occupied with an
electron of spin $-\sigma$. The exchange energy for this and the 
opposite process is
\begin{equation}
J_{ij} = 
\frac{t^2}{U-(\epsilon_i-\epsilon_j)}+\frac{t^2}{U-(\epsilon_j-\epsilon_i)}
\approx \frac{2t^2}{U} \left[1+\frac{(\epsilon_i-\epsilon_j)^2}{U^2}\right] .
\end{equation}
$J_{ij}$ is hence always \em enhanced \em from the pure case $(J_0=2t^2/U)$,
and the average exchange becomes $J_{\rm av}=J_0 [1+\lambda(\Delta/U)^2]$
where $\lambda$ depends only on the shape of the disorder distribution
(e.g., $\lambda=1/12$ for the constant, $\lambda=1/2$ for a binary
 distribution). 
Furthermore, one can show that the form of the magnon propagator remains 
unchanged in second order in $\Delta/U$ \cite{singh98}, but magnons
are stiffened by a factor of $(1+\lambda(\Delta/U)^2)$.
The effective hopping in strong coupling can thus 
be expressed as $t_{\rm eff}= t [1+\lambda(\Delta/U)^2]^{1/2}$.
In dimensions $d>2$ one expects the N\'eel temperature $T_N$ to be proportional
to $J$ and hence $T_N$, too, should be enhanced by weak disorder.

While disorder-enhanced delocalization 
stabilizes AF order at strong coupling it suppresses it
at weak coupling. For small $U$ the AF state is rather ``spin density
wave like'' and an enhanced
kinetic energy tends to weaken AF order and to reduce $T_N$.
In addition, random potentials destroy the perfect nesting instability
responsible for AF order at small $U$.

The $T$ vs.~$U$ phase diagram within DMFT (Fig.~3) summarizes and confirms
the above considerations: For small $U$, $T_N$ is reduced and eventually
vanishes when $\Delta$ becomes roughly equal to $U$. At large $U$, however,
the $T_N$ curves for different $\Delta$ cross each other, i.e.~at
a given value of $U$, $T_N$ increases
with disorder. 
The opposite effects of disorder on $T_N$ depending on $U$ are due to
the non-monotonic behavior of the function $T_N(U)$. 
If the slope of $T_N(U)$ is positive (negative) an effectively
reduced ratio $U/t_{\rm eff}$ leads to a suppression (enhancement) of $T_N$.

\subsection{Closing of the charge gap}

Local potentials can induce electronic states in the charge gap.
As also found experimentally the positions of the impurity states
are crucial for the stability of AF order with respect to carrier
doping. 
Disorder-induced gap states will reduce the size of the charge gap. 
We now want to determine the critical disorder strength at which the gap
vanishes.
First we treat the disorder within the T-matrix approximation
which becomes exact for a single impurity 
\cite{sen,singh98}.
Since the local host Green function of the correlated, not-disordered problem
is unknown we approximate it by the 
Green function of the AF Hartree-Fock solution:
\begin{equation}
[g^{0}_{\sigma}]_{ii}=(1/N)\sum_{\bf k} (\omega- \sigma D)/
(\omega^2 - E_{\bf k} ^2) .
\end{equation}
Here $2D=mU$ is the Hubbard energy gap in the pure AF,
$E_{\bf k}=\sqrt{D^2 +\epsilon_{\bf k}^2}$ is the AF band energy, 
and $D$ is obtained from the self-consistency condition
$(1/N)\sum_{\bf k} (2E_{\bf k})^{-1} = U^{-1}$. 
With this local host Green function we can calculate the location of 
impurity-induced states from the poles in the T-matrix,
$T_{\sigma}(\omega)=
\epsilon_i/(1-\epsilon_i[g^{0}_{\sigma}(\omega)]_{ii})$.
We consider a constant distribution of random potentials
between $\pm\Delta/2$. There are disorder-induced states within 
the gap if $|[g^{0}_{\sigma}]_{ii}|>2/\Delta$.
The energy $\tilde D$ up to which states are formed within the gap is
given by $[g^0_{\sigma}(-\tilde{D})]_{ii}= 2/\Delta$.
The remaining charge gap $2\tilde D$ is plotted versus $\Delta$ in Fig.~4.
The decrease is almost linear and $2\tilde D$ vanishes close to
$\Delta=U$. Also shown is the result from a numerical unrestricted 
Hartree-Fock (UHF) analysis. In this approach, the HF Hamiltonian on a 
finite lattice is numerically (self-consistently) diagonalized,
so that disorder is treated exactly \cite{logan,uhf,singh98}.
The energy gap is obtained from the 
energy difference between the lowest energy state of the
upper Hubbard band and the highest energy state of the lower Hubbard band.
Averages are taking over 100 disorder realizations
on a $10\times 10$ lattice.
The saturation of the gap at $\Delta/U\sim 1$ is due to the finite system size.
The agreement with the T-matrix approach is excellent for the present 
interaction value $U=10t$.
Deviations from the T-matrix approach 
are more pronounced at lower interaction strengths where 
the fermion states are more extended. 

\begin{figure}
\vspace{-10mm}
\psfig{file=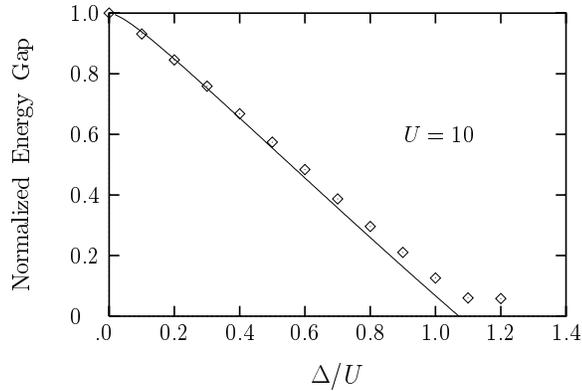,height=0.53\hsize}
\caption{
Normalized energy gap, $\tilde D/D$, in $d=2$ versus disorder strength 
$\Delta$ at $U=10t$
as obtained by the T-matrix approximation (solid line) 
and the UHF analysis (symbols) \cite{singh98}.
}
\end{figure}
\begin{figure}
\vspace{-5mm}
\psfig{file=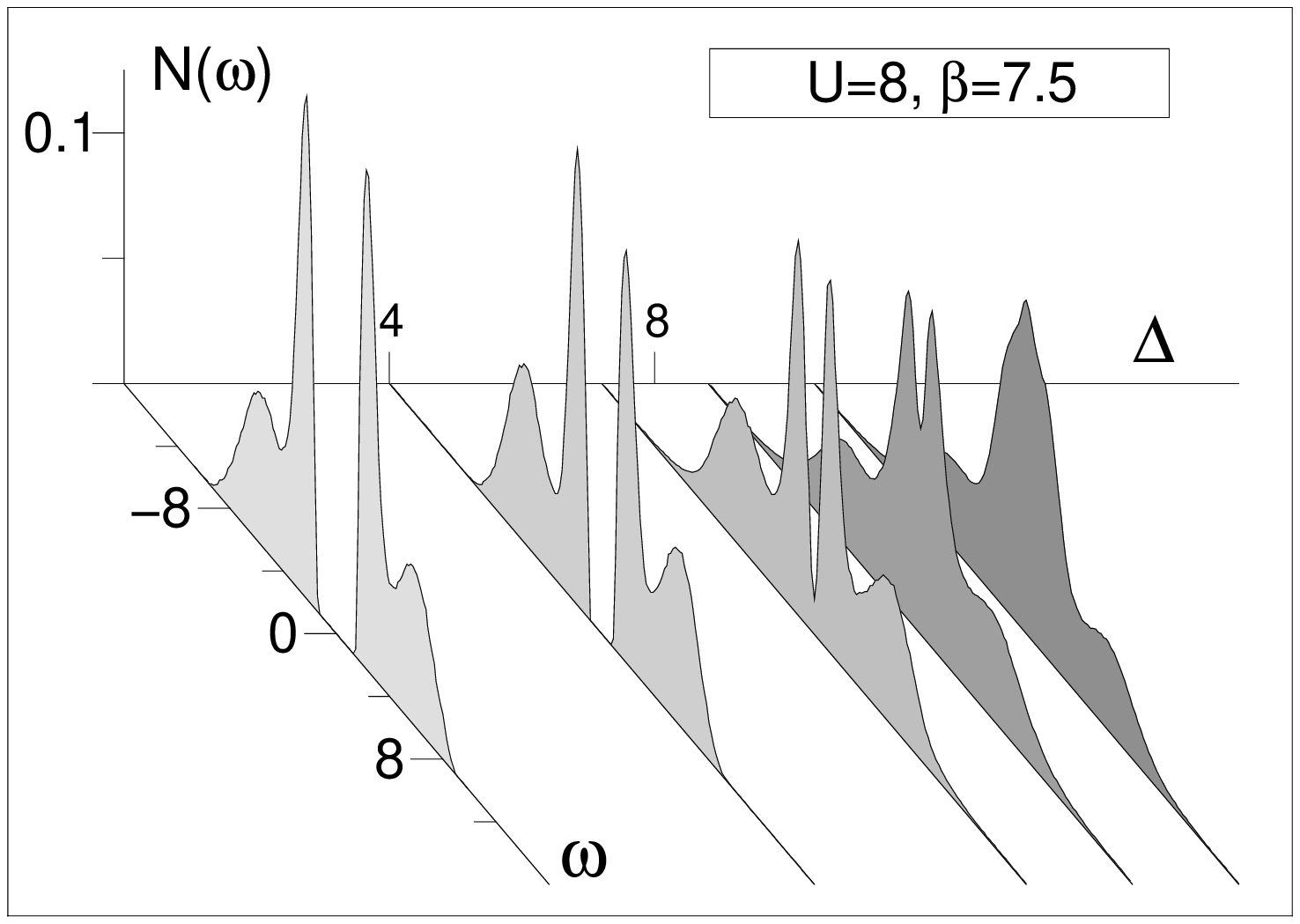,height=0.5\hsize}
\vspace*{-62mm}
\hspace*{60mm}
\psfig{file=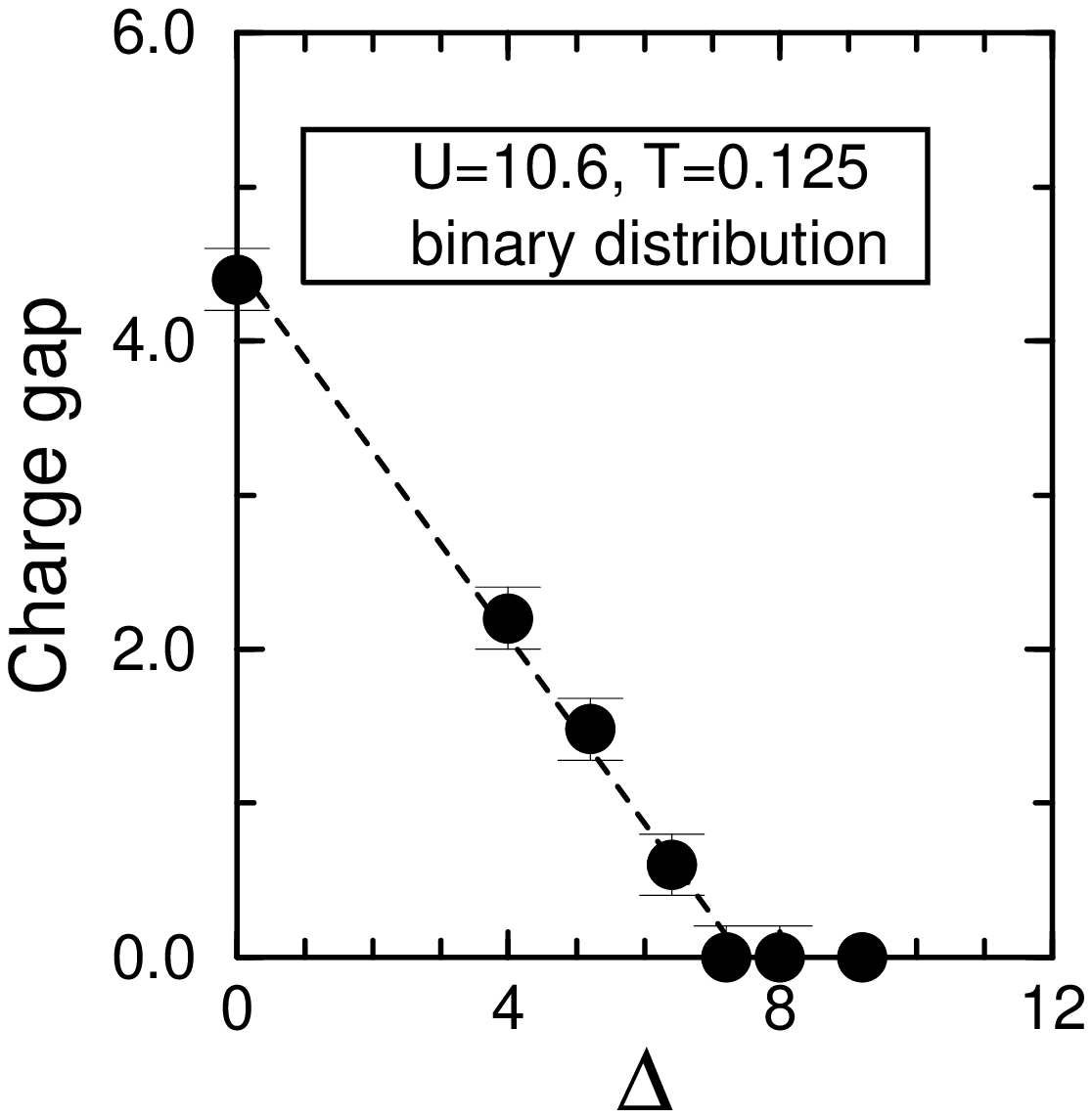,height=0.54\hsize}
\vspace*{-15mm}
    \par\makebox[0.4\hsize]{\small (a)}\hspace{\fill}%
    \makebox[0.55\hsize]{\small (b)}
\caption{
(a) Density of states (DOS) for a constant disorder distribution 
of different width in DMFT \cite{singh98};
all spectra are within the AF phase. (b) Charge gap vs.~$\Delta$ in DMFT for
the bimodal distribution of random potentials.
}
\end{figure}

The closing of the charge gap is also observed within the
DMFT approach. Fig.~5a shows the density of states for several
disorder values. Note that all spectra shown are within the
AF ordered phase, i.e.~AF order is much more stable than the charge gap
which closes about $\Delta=U$.
A linear reduction of the charge gap is also observed in the case
of a binary disorder distribution (Fig.~5b). For this stronger type of 
disorder both the charge gap and the AF order vanish near $\Delta=U$.

\subsection{Spin vacancies}

As discussed in the previous section the one-particle excitation gap 
decreases with $\Delta$ and vanishes near $\Delta=U$.
Upon further increase of $\Delta$ the two Hubbard bands will overlap
whereby electrons from the highest levels of the 
lower band (with $\epsilon_i>U/2$) will be transfered to the lowest
levels of the upper band (with $\epsilon_i<-U/2$). Those sites
will become doubly occupied, their local moment will vanish as observed
in Sec.~\ref{Sec:localmoments}.
In the limit $t\ll U,\Delta$ the Hubbard bands at $\pm U/2$
are nearly flat with width $\Delta$. 
The overlap region is $\sim(\Delta-U)$ and the fraction of nonmagnetic
sites can be estimated by $x\sim (\Delta-U)/\Delta$.

In the following
this situation is modeled by spin vacancies of concentration $x$
which will lead to strong magnon scattering.
Such diluted antiferromagnets have been widely studied recently
mostly within spin diluted Heisenberg models 
\cite{bulut,kampf} and also
Hubbard models \cite{sen}.
A perturbative analysis in the strong coupling limit $({\cal O} (t^2/U))$
\cite{sen,singh98} yields a softening of magnons by a factor of $(1-2x)$ in
$d=2$. Extrapolating to large fractions $x$, the magnon energy scale 
$\tilde J = J (1-2x)$ hence vanishes only close to the percolation 
threshold $x_{\rm perc}\approx0.4$.
In $d>2$ the N\'eel temperature is assumed to be proportional to $\tilde J$
and a linear decrease of $T_N$ is thus expected.
In experiments, e.g.~on Li doped NiO \cite{Corti} 
such a linear dependence on $x$ is 
indeed observed with a prefactor of 2.2.
\begin{figure}[b]
\vspace*{-15mm}
\psfig{file=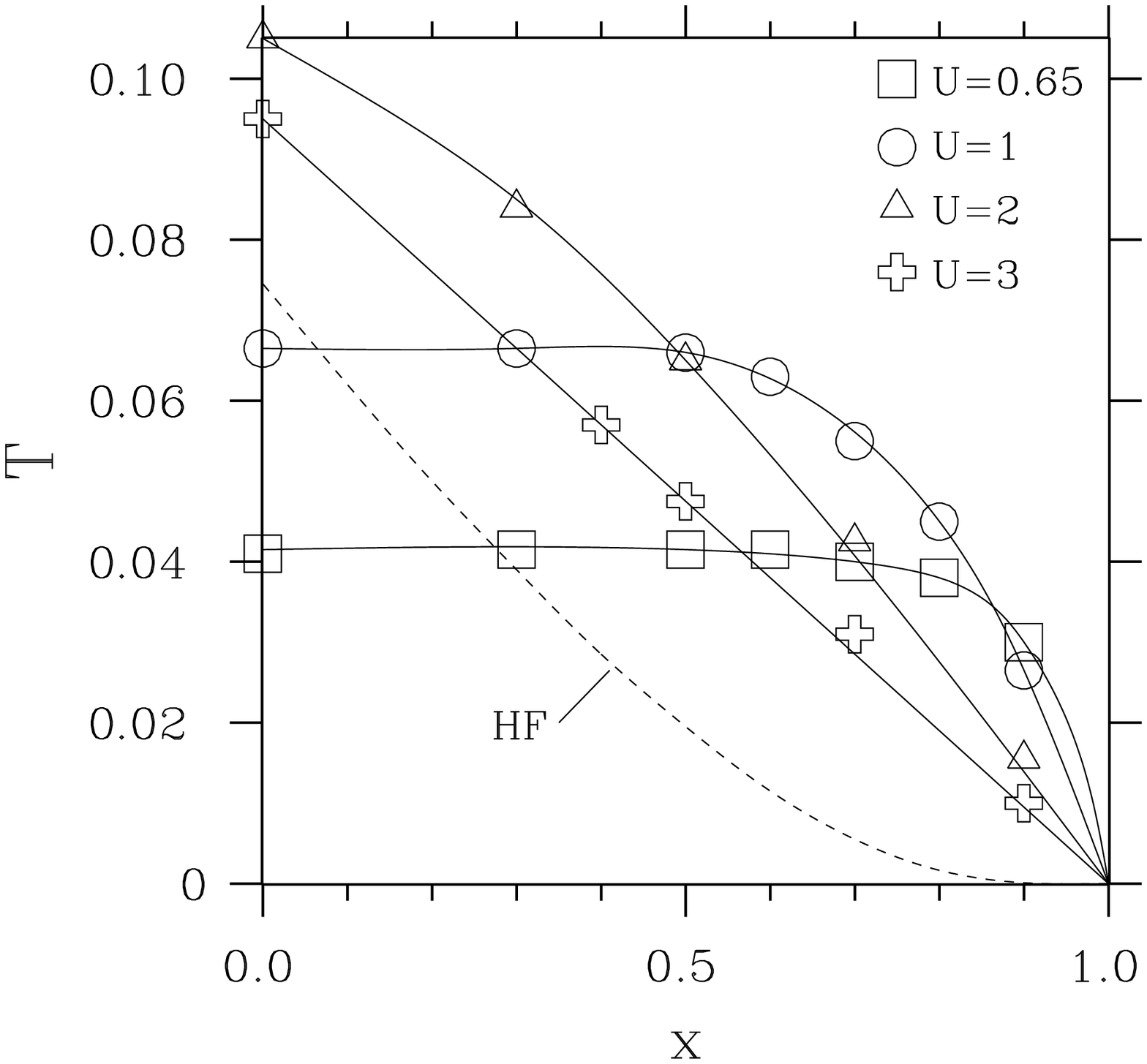,height=0.6\hsize}
\vspace*{-78mm}
\hspace*{50mm}
\psfig{file=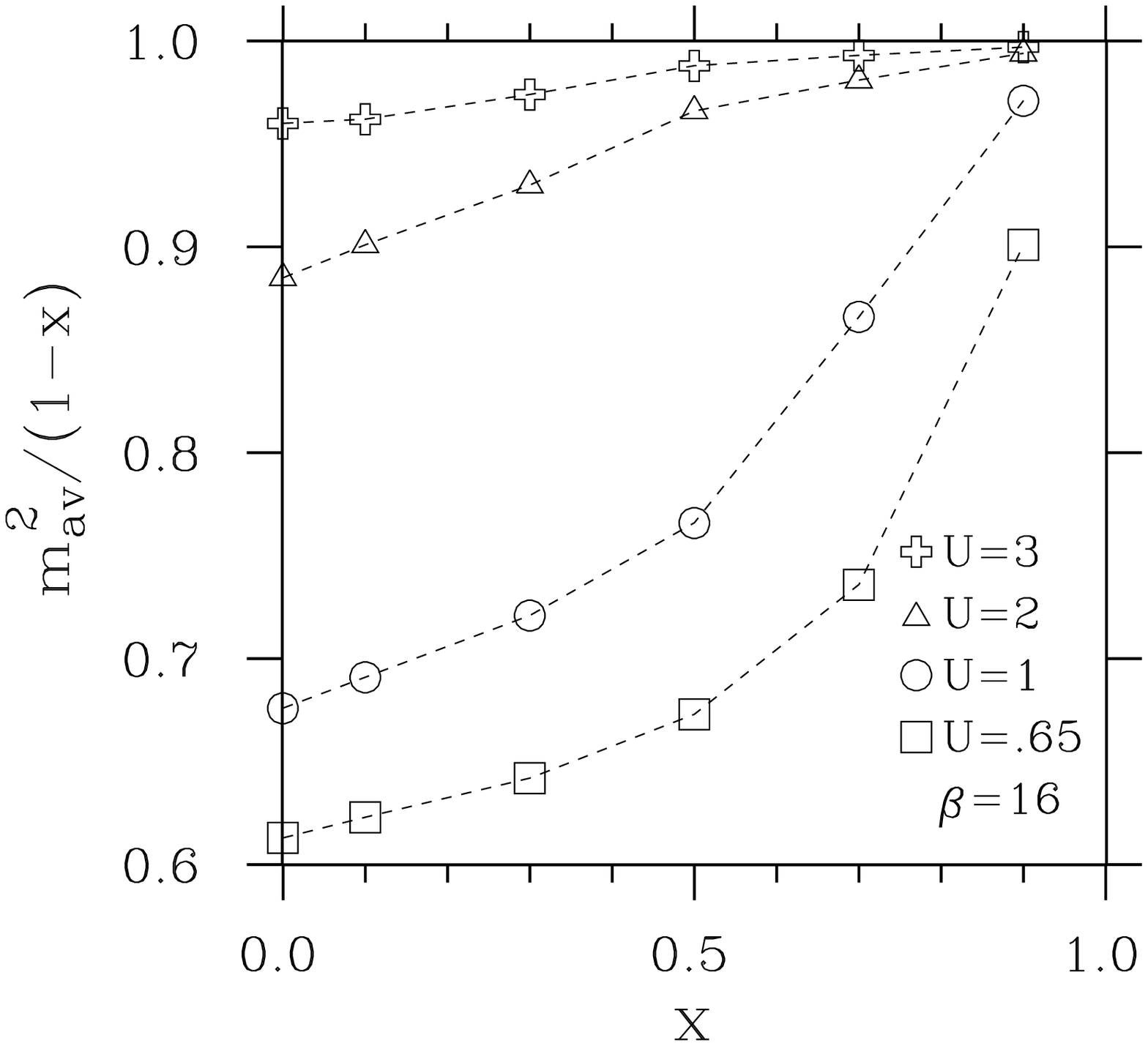,height=0.669\hsize}
\vspace*{-15mm}
    \par\makebox[0.4\hsize]{\small (a)}\hspace{\fill}%
    \makebox[0.55\hsize]{\small (b)}
\caption{
(a) $T_N$ as a function of site dilution in DMFT for different values of $U$.
Dashed line Hartree-Fock approximation for $U=0.65$ 
(energies in units of the half bandwidth).
(b) Enhancement of local moments on the remaining fraction $(1-x)$
of the lattice at inverse temperature $\beta=16$. \cite{ulmke}
}
\end{figure}
In the limit $d\to\infty$ the percolation threshold is 1.0; hence
we can expect AF order to persist for arbitrarily large dilution.
Within DMFT, too, a linear decrease of $T_N\propto (1-x)$ is observed for 
the diluted model at large $U$ (Fig.~6a). At small $U$, however,
AF order is remarkably robust, $T_N$ being constant up to dilutions
close to $x=1$ where $T_N$ eventually drops to zero.
This behavior of $T_N$ can be explained by a strong enhancement
of the local moment density on the remaining sites at small $U$ (Fig.~6b).
The reason for this enhancement is that with reduced average 
number of nearest neighbors
the kinetic energy decreases, leading to a stronger localization.
At large $U$ the local moments are already almost saturated
at $x=0$ and just cannot be further enhanced.
The situation in the case of vacancies is therefore quite different from the
effect of weak site disorder where disorder-enhanced \em delocalization \em
is observed.

\section{Random hopping}

The case of spin vacancies discussed in the previous section
can of course be regarded as a specific type of randomness
in the hopping elements. The more generic case of a continuous
(flat) distribution of $t_{ij}\in [1-\Delta/2,1+\Delta/2]$
was studied in $d=2$ using QMC.
Since the hopping is still restricted between nearest neighbors
on a square lattice particle-hole symmetry is preserved and no
minus-sign problem occurs at $n=1$.
Random $t_{ij}$ hardly affect the density of local moments
(Fig.~7a), the slight decrease may be due to the enhanced kinetic
energy which is $\propto \sqrt{<t_{ij}^2>}$.
Nevertheless, longer-range spin-spin correlations are strongly suppressed
if $\Delta$ is of the order of $t$ (see Fig.~7a).
The finite size scaling according to (\ref{Eq:fss}) yields the 
AF order parameter (staggered moment) $M$ versus $\Delta$ (Fig.~7b).
$M$ vanishes at a critical disorder strength of $\Delta_c\approx 1.4$.
We propose \cite{ulmke2} that the phase boundary is determined basically
by the variance of the AF exchange coupling 
$v=(<J_{ij}^2>-<J_{ij}>^2)/<J_{ij}^2>$. AF order persists for
$v<v_c\approx 0.4$. This criterion is consistent with the phase boundary
of the bond-disordered AF Heisenberg model with a bimodal 
distribution of $J_{ij}$ \cite{sandvik2}.
The reason for the vanishing of AF order for this type of disorder
is supposedly the formation of local singlets. Such singlets will
form first on the strongest bonds and will leave some spins which are
weakly coupled to their neighbors unpaired. Those ``free'' spins are expected
to give a Curie-like contribution to the susceptibility as observed
in doped semiconductors \cite{belitz,bhatt}.
The numerical results indeed show a strong enhancement of the uniform
susceptibility in the disordered case \cite{ulmke2}. 

\begin{figure}
\vspace*{-5mm}
\psfig{file=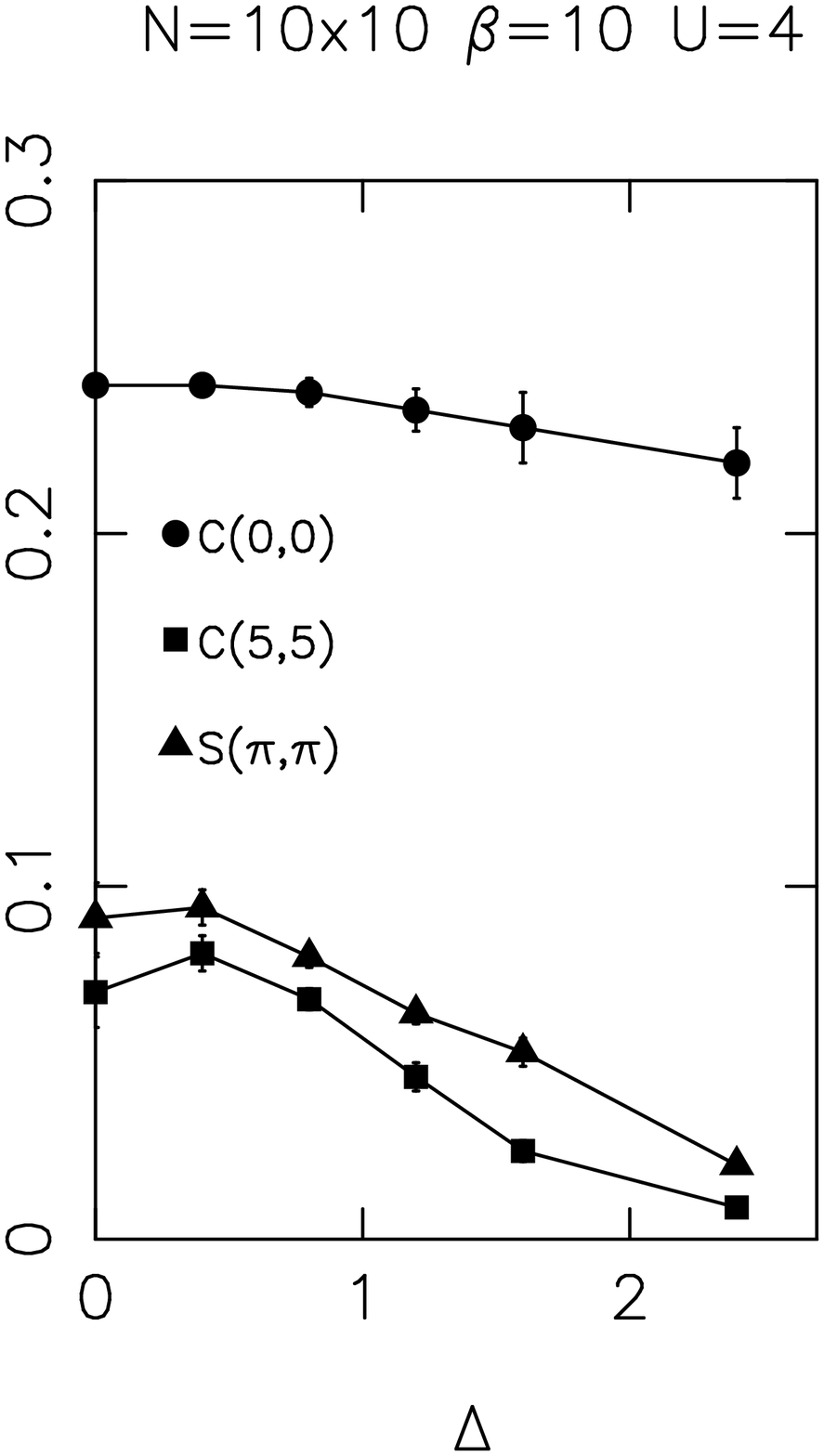,height=0.65\hsize,width=0.6\hsize}
\vspace*{-79.7mm}
\hspace*{60mm}
\psfig{file=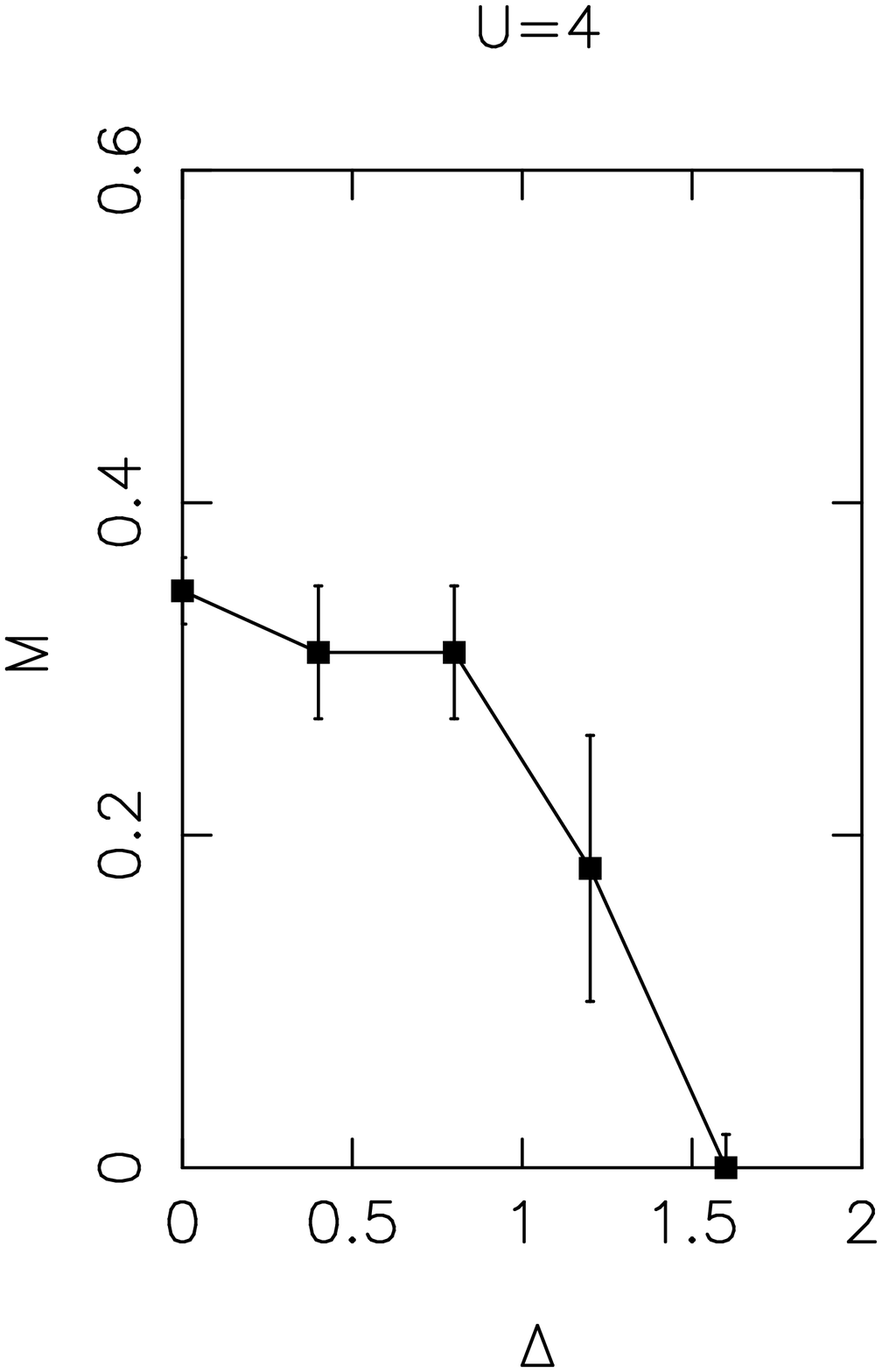,height=0.65\hsize,width=0.6\hsize}
\vspace*{-10mm}
    \par\makebox[0.4\hsize]{\small (a)}\hspace{\fill}%
    \makebox[0.55\hsize]{\small (b)}
\caption{(a) Local moment and spin-spin correlations in $d=2$ for
random hopping (same quantities as in Fig.~1).
(b) Staggered magnetization $M$ vs.~$\Delta$ 
as obtained by finite size scaling \cite{ulmke2}.
}
\end{figure}

\section{Impurities with weak local interactions}

The Hubbard model at $n=1$ exhibits both AF order and the 
Mott-Hubbard metal-insulator transition (MIT).
It is important to note that both effects are in principle
independent. The Mott-Hubbard MIT occurs at intermediate
interaction $(U\sim \rm bandwidth)$ while AF order can set in
at arbitrarily small $U$ due to the Fermi surface nesting instability.
It is argued that the MIT at $T=0$ represents a
quantum critical point which is however concealed by the 
low temperature AF phase \cite{gebhard}, and attempts are made
to suppress the AF phase to very low or zero $T$
by different types of frustration \cite{pruschke}.
A different possibility to separate AF order and the MIT is to
shift the filling at which the Mott-Hubbard MIT occurs away
from $n=1$ by introducing impurities with a low (or zero) local 
interaction with a concentration $f$.
Such impurity sites can be doubly occupied without the cost
of the local repulsion and hence the MIT is expected at 
a density $n=1+f$.

\subsection{Mott-Hubbard gap}
The Hubbard model with a bimodal distribution of $U$ values
($U=0$ on a fraction $f$ of the lattice, and $U=8t$ on the remaining sites)
was investigated by QMC on square lattices and in DMFT.
Note that for technical reasons (minus-sign problem) the 
choice of the $U-$disorder in model (\ref{Eq:dishub})
preserves particle-hole symmetry.
This corresponds to different chemical potentials on the two
constituents such that at $n=1$ the electronic density is homogeneous,
i.e.~independent of $U_i$.
To detect charge gaps the electron density $n$ is plotted versus the
chemical potential $\mu$ (Fig.~8a). As expected the gap moves
to a density off half filling close to $1+f$. This agrees with 
the kinetic energy which shows minima at the corresponding densities
(Fig.~8b). At a closer look one observes an additional gap at
$n=1$ which is due to the doubling of the unit cell in the AF ordered state.
The gaps at $n=1$ and $n=1+f$ can also be detected by the one-particle
spectrum, obtained in DMFT (see \cite{ulmke3}). 
Between densities 1 and $1+f$ the systems is apparently compressible. 

\begin{figure}
\psfig{file=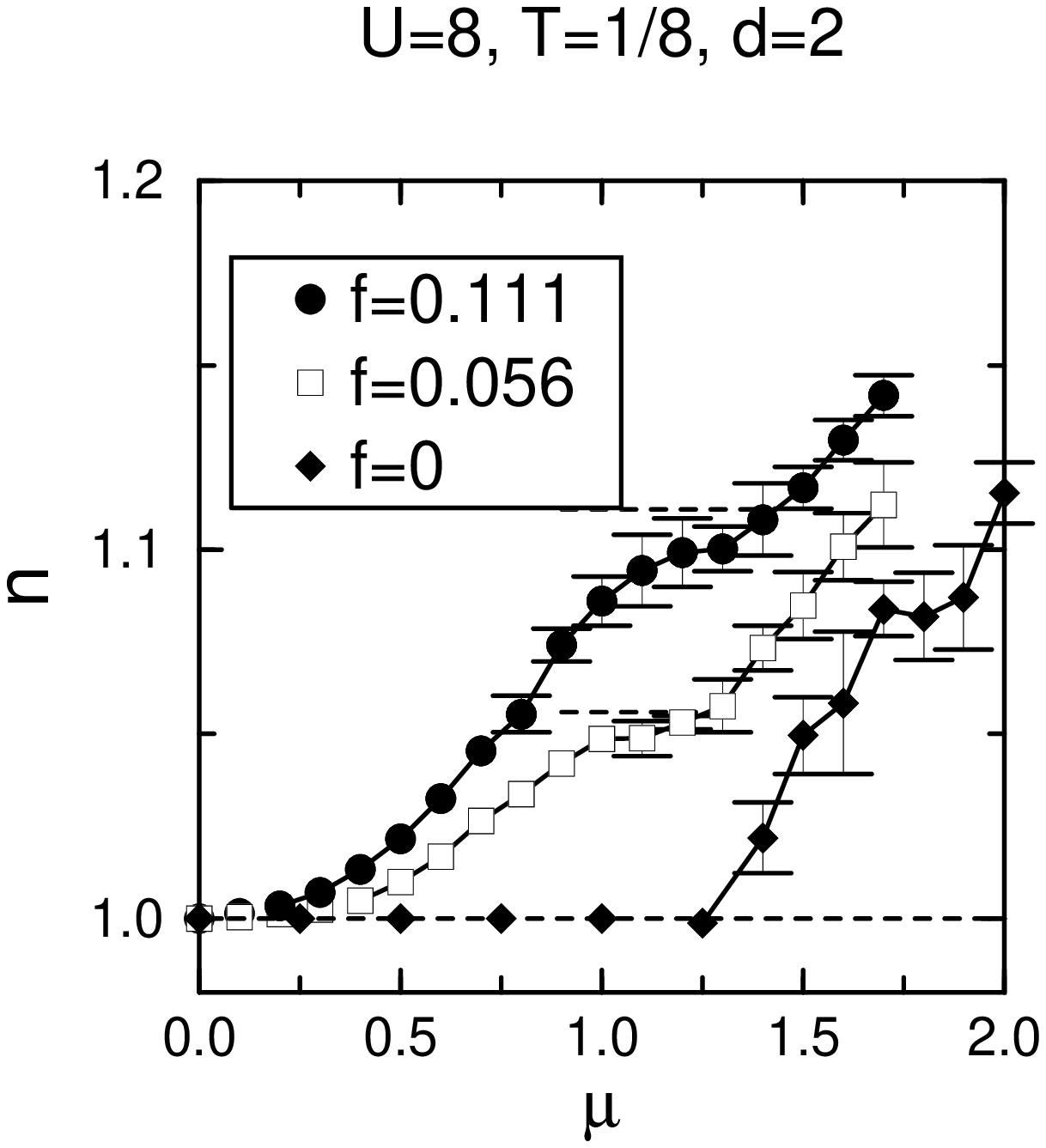,height=0.6\hsize}
\vspace*{-89.5mm}
\hspace*{47mm}
\psfig{file=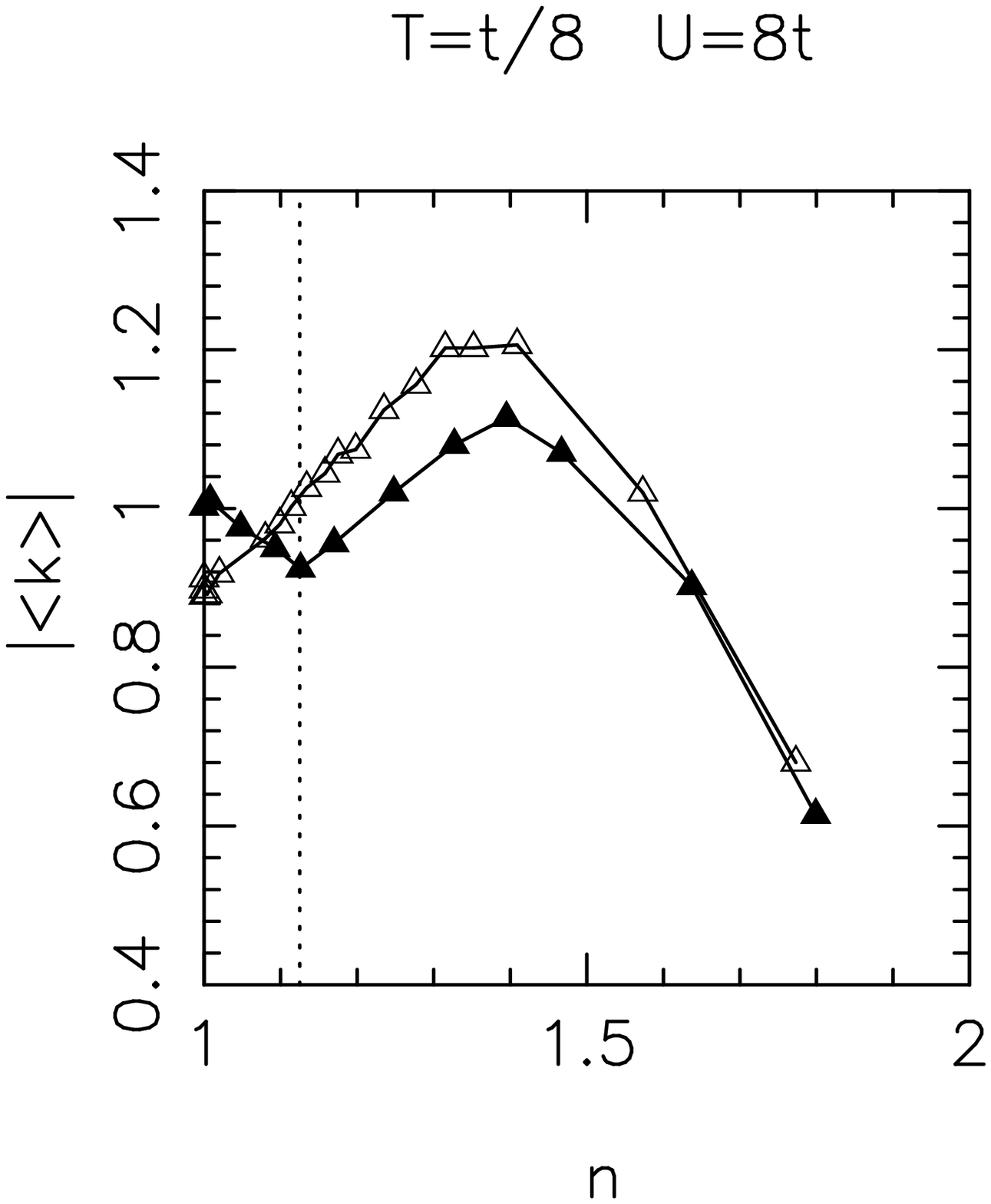,height=0.7036\hsize}
\vspace*{-10mm}
    \par\makebox[0.4\hsize]{\small (a)}\hspace{\fill}%
    \makebox[0.55\hsize]{\small (b)}
\caption{(a)
Electron density $n$ vs.~chemical potential $\mu$ for different values
of $U=0$ impurity concentration $f$. $n(\mu)$ has plateaux at $n=1$ and close
to $n=1+f$.
(b)
Kinetic energy $\langle k\rangle$ vs.~average density for 
 $f=0$ (open triangles) and  $f=0.111$ (solid triangles).
$\langle k\rangle$ shows minima at $n=1+f$ \cite{ulmke3}.
}
\end{figure}

\subsection{Antiferromagnetic order}

Figure 9a shows the staggered moment $M$ 
extrapolated to the thermodynamic limit in $d=2$ at $n=1$
as a function of impurity concentration $f$. 
For small $f$ AF order is robust leading to the charge gap at $n=1$.
AF order vanishes at $f_c\approx 0.45$,
i.e.~close to the percolation threshold $f_{\rm perc}\approx 0.4$.

Off half-filling the finite size scaling is no longer possible
due to the minus-sign problem and the AF phase boundary is obtained
within DMFT only (shown in Fig.~9b at $T=1/8$).
It is found that the $U=0$ impurities can induce AF order at densities
for which the clean model is disordered.
In the clean model $(f=0)$ additional electrons are free to move
and hence very effective in destroying long range order.
$U=0$ sites provide localizing centers which are energetically
favorable for the additional carriers. This is why the AF phase
extends to larger dopings at finite (but small) values of $f$.
Since the mechanism of localizing mobile dopants is observed in $d=2$, too,
we expect the enlargement of the AF phase to be present in $d=2$ 
at $T=0$ as well,
in spite of the fact that in the clean model in $d=2$ the critical 
doping is supposed to be zero.
Eventually at larger fractions $f$ the AF phase shrinks, and the
critical density approaches 1.0 for $f_c\approx 0.6$ 
($f_c\approx 0.75$ in the ground state \cite{ulmke3}). For even larger
values of $f$ there is no AF order even at $n=1$.

Both the stabilization of AF order and the shift of the Mott gap
to higher densities result from the localization of carriers at the
$U=0$ impurities. AF order is only destroyed when the density on the
$U=0$ sites saturates, i.e.~it is stable \em across \em the Mott-Hubbard
MIT. Hence the separation of Mott-Hubbard MIT and AF order is not
present in the $U=0$ impurity model, at least not for the present 
parameter values.

\begin{figure}
\vspace*{-15mm}
\psfig{file=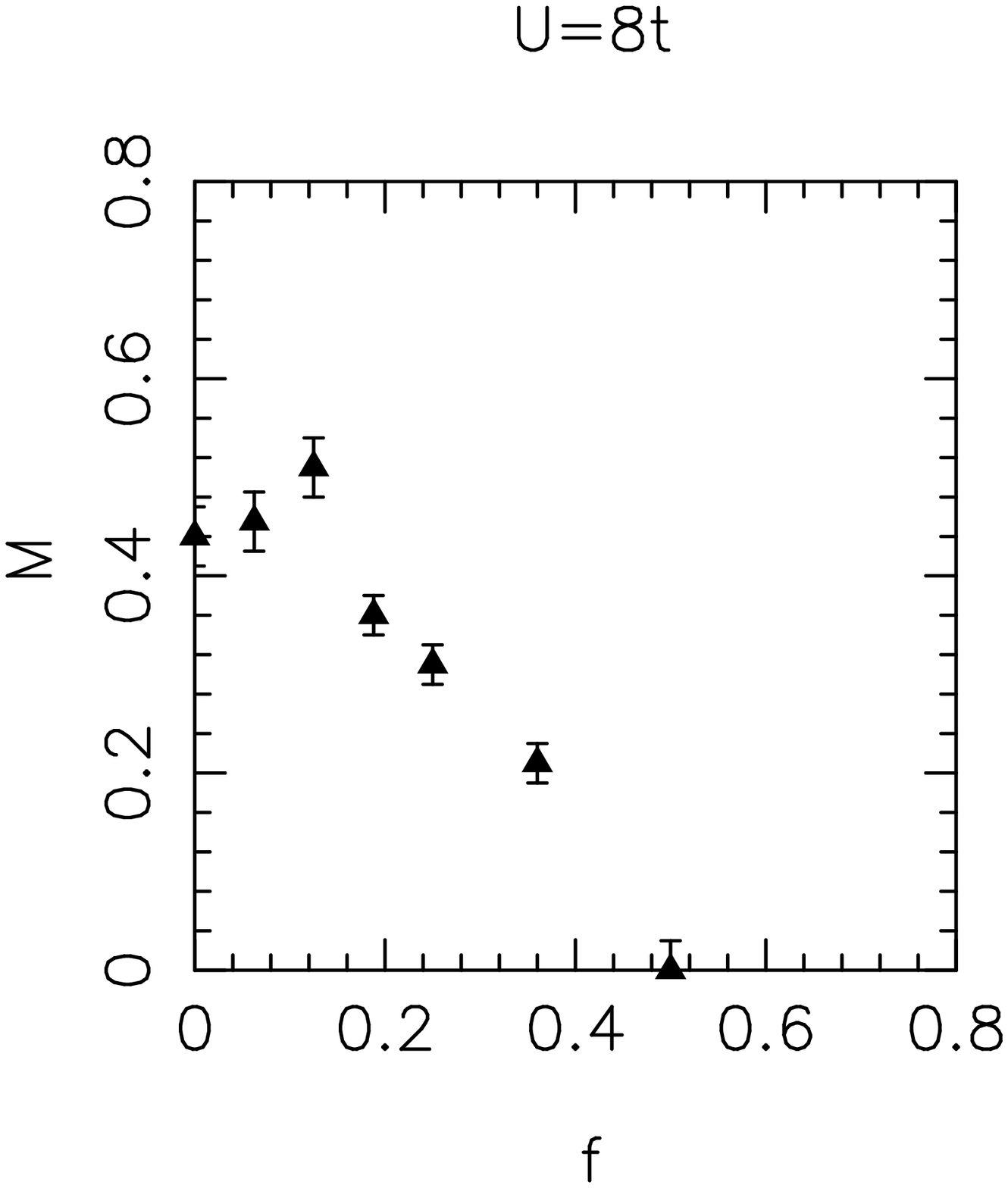,height=0.68\hsize}
\vspace*{-70mm}
\hspace*{55mm}
\psfig{file=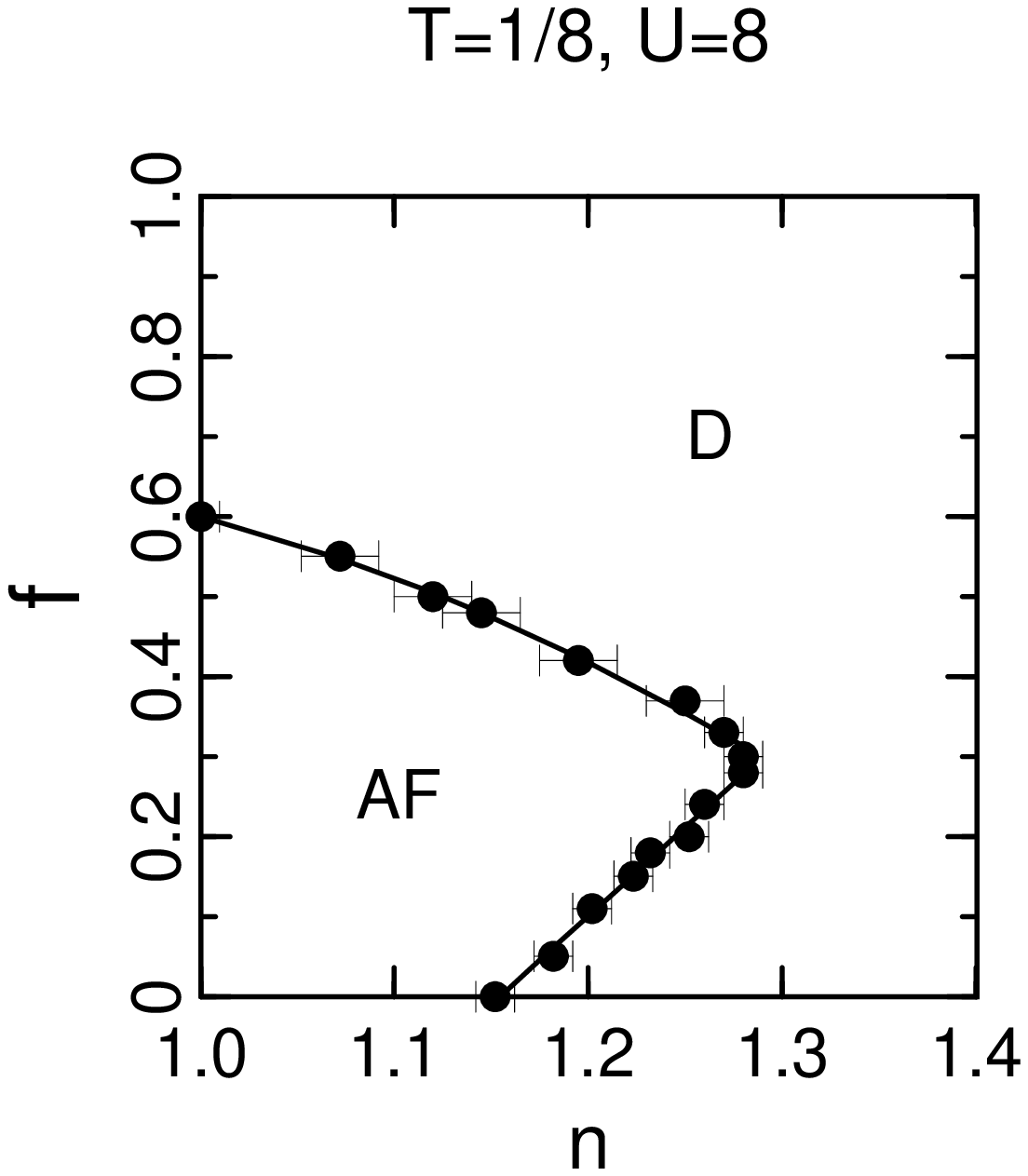,height=0.6\hsize}
\vspace*{-15mm}
    \par\makebox[0.4\hsize]{\small (a)}\hspace{\fill}%
    \makebox[0.55\hsize]{\small (b)}
\caption{
(a) Staggered magnetization in $d=2$ as a function of impurity 
concentration $f$.
(b) $f-n$ phase diagram at temperature $T=1/8$ within DMFT.
For small $f$ the AF phase is stabilized against doping. 
\cite{ulmke3}}
\end{figure}

\section{Summary and Conclusion}

In this paper we discussed different 
types of disorder in correlated antiferromagnets
and presented results obtained mostly by quantum
Monte Carlo simulations in $d=2$ and within dynamical
mean-field theory ($d=\infty$).

Different mechanisms were identified by which disorder
can enhance antiferromagnetic order: i) disorder-enhanced
delocalization at strong coupling in the case of weak disorder 
in the chemical potentials
and ii) localization of surplus carriers in the case of
impurities with weak local interaction.
In both cases compressible antiferromagnetic phases are observed.
To determine if the gapless AF phase is metallic requires the calculation
of transport properties which is presently in progress.

Quantum Monte Carlo simulations of electronic tight binding 
models are just making the transition from addressing 
rather abstract issues of correlation effects to making 
contact with real experiments.
One important feature in this respect is the treatment
of intrinsic disorder. 
For a quantitative description of experiments, however, 
the inclusion of a realistic bandstructure is mandatory. 
Here the DMFT will be particularly helpful because
it allows for the treatment of multiband models
in a wider range of parameters and in the thermodynamic limit.


\end{document}